\documentclass[aps,amsmath,amssymb,prb,10pt,twocolumn,superscriptaddress]{revtex4-1}
\usepackage{graphicx,txfonts}
\usepackage{bm}
\usepackage{color}
\usepackage{mathrsfs}
\usepackage{comment}
\usepackage{mathtools}
\mathtoolsset{showonlyrefs}

\newcommand{\al}{\alpha}
\newcommand{\be}{\beta}
\newcommand{\g}{\gamma}
\newcommand{\de}{\delta}

\newcommand{\z}{\zeta}

\newcommand{\thi}{\theta}

\newcommand{\ka}{\kappa}
\newcommand{\la}{\lambda}

\newcommand{\p}{\pi}

\newcommand{\f}{\phi}
\newcommand{\x}{\chi}
\newcommand{\w}{\omega}
\newcommand{\W}{\Omega}
\newcommand{\De}{\Delta}
\newcommand{\G}{\Gamma}

\newcommand{\pd}{\partial}

\newcommand{\round}[1]{\left( #1 \right)}
\renewcommand{\square}[1]{\left[ #1 \right]}

\newcommand{\ang}[1]{\left\langle #1 \right\rangle}

\newcommand{\bra}[1]{\left\langle #1\right|}
\newcommand{\ket}[1]{\left| #1\right\rangle}

\newcommand{\beq}{\begin{equation}}
\newcommand{\eeq}{\end{equation}}
\newcommand{\Beq}{\begin{eqnarray}}
\newcommand{\Eeq}{\end{eqnarray}}
\newcommand{\bml}{\begin{multline}}

\newcommand{\bea}{\begin{align}}
\newcommand{\ena}{\end{align}}
\newcommand{\bsp}{\begin{split}}
\newcommand{\esp}{\end{split}}

\newcommand{\bM}{{\boldsymbol M}}

\newcommand{\ey}{\hat{\boldsymbol y}}
\newcommand{\ez}{\hat{\boldsymbol z}}

\renewcommand{\bm}{{\boldsymbol m}}

\newcommand{\bF}{{\boldsymbol F}}

\newcommand{\bB}{{\boldsymbol B}}
\newcommand{\bb}{{\boldsymbol b}}

\DeclareMathOperator{\diag}{Diag}

\newcommand{\sA}{\mathscr{A}}

\newcommand{\hH}{\hat{H}}

\newcommand{\bH}{\boldsymbol{H}}

\newcommand{\bpi}{\boldsymbol{\pi}}

\newcommand{\bx}{\boldsymbol{x}}
\newcommand{\bn}{\boldsymbol{n}}

\newcommand{\bs}{{\boldsymbol{s}}}

\newcommand{\hU}{\hat{U}}

\newcommand{\hA}{\hat{A}}

\newcommand{\hsig}{\hat{\sigma}}

\begin{document}
\title{Quantum control of topological defects in magnetic systems}
\author{So Takei}
\affiliation{Department of Physics, Queens College of the City University of New York, Queens, NY 11367, USA}
\affiliation{The Physics Program, The Graduate Center of the City University of New York, New York, NY 10016, USA}
\author{Masoud Mohseni}
\affiliation{Google Inc., Venice, CA 90291, USA}
\date{\today}

\begin{abstract}
Energy-efficient classical information processing and storage based on
topological defects in magnetic systems have been studied over past decade. In this work, we introduce a class of macroscopic quantum devices in which a quantum state is stored in a topological defect of a magnetic insulator. We propose non-invasive methods to coherently control and readout the quantum state using ac magnetic fields and magnetic force microscopy, respectively. This macroscopic quantum spintronic device realizes the magnetic analog of the three-level rf-SQUID qubit and is built fully out of electrical insulators with no mobile electrons, thus eliminating decoherence due to the coupling of the quantum variable to an electronic continuum and energy dissipation due to Joule heating. For a domain wall sizes of $10-100$~nm and reasonable material parameters, we estimate qubit operating temperatures in the range of $0.1-1$~K, a decoherence time of about $0.01-1$~$\mu$s, and the number of Rabi flops within the coherence time scale in the range of $10^{2}-10^{4}$. \end{abstract}
\maketitle

\section{Introduction}
Topological spin structures are stable magnetic configurations that can neither be created nor destroyed by any continuous transformation of their spin texture.~\cite{merminRMP79} Its topological robustness, small lateral size, and relatively high motional response to electrical currents have inspired proposals utilizing these topological defects as bits of information in future data storage and logic devices.~\cite{kiselevJPD11,omariPRAP14} Magnetic skyrmions (vortex-like topological spin structures)~\cite{rosslerNAT06} have defined the field of skyrmionics where every bit of information is represented by the presence or absence of a single skyrmion~\cite{kiselevJPD11,rommingSCI13,malozemoffBOOK16}. Recent proposals have exploited their miniature size and high mobility for skyrmion-based logic gates~\cite{zhangSR15lg}, racetrack memories~\cite{fertNATN13}, high-density non-volatile memory~\cite{koshibaeJJAP15}, energy-efficient artificial synaptic devices in neuromorphic computing~\cite{huangNT17} as well as reservoir computing platform~\cite{prychynenkoCM17}. More recently, storage and processing of digital information in magnetic domain walls and their potential uses for logic operations have been discussed.~\cite{nakataniJMMM05,omariPRAP14,vandermeulenJPD15,malozemoffBOOK16} Although research to exploit topological magnetic defects for energy-efficient classical information technology are being actively pursued, the study of how quantum information can be stored, manipulated and readout in such defects still remains unexplored.

Macroscopic quantum phenomena in magnetic systems have been studied over the past three
decades.~\cite{chudnovskyPRL88,*barbaraPLA90,*awschalomPRL92,*giderSCI95,*guntherBOOK95,*chudnovskyBOOK98} Recently
we have proposed a spin-based macroscopic quantum device that is realizable and controllable with current spintronics
technologies.~\cite{takeiPRB17} The work proposed a ``magnetic phase
qubit", which stores quantum information in the orientational degree
of freedom of a macroscopic magnetic order parameter and offers the
first alternative qubit of the macroscopic type to superconducting
qubits. In principle, the upper bound for the operational temperature
of a magnetic quantum device is set by magnetic ordering temperatures,
which are typically much higher than the critical temperatures of
superconducting Josephson devices.  Estimates based on the
state-of-the-art spintronics materials and technologies led to an
operational temperature for the magnetic phase qubit that is more than
an order of magnitude higher than the existing superconducting qubits,
thus opening the possibility of macroscopic quantum information
processing at temperatures above the dilution fridge range. The
proposed qubit is controlled and readout electrically via spin Hall
phenomena~\cite{sinovaRMP15}, which involves a coupling of the qubit
to a current-carrying metallic reservoir. Based
on the current spin Hall heterostructures, significantly high dc
charge current densities in the metal 
may be necessary for qubit operation. Therefore, a challenging aspect of the proposal is how to remove the excess heat that is generated by this control current. It is also desirable to propose an alternative macroscopic magnetic qubit that is fully operational without electrical currents and precludes Joule heating.

In this work, we propose how a macroscopic quantum state can be stored, manipulated and readout using a soft collective mode~\cite{tretiakovPRL08} of a topological defect, i.e., a domain wall, in an insulating magnetic material, and non-invasive control and readout of this state using ac magnetic fields and magnetic force microscopy, respectively. As the soft mode collectively describes a macroscopic number of microscopic spins in the magnet, such a device is a macroscopic quantum device, and can be built, in principle, using existing solid-state technologies. The device realizes the magnetic analog of a three-level rf-SQUID qubit that is built fully out of electrical insulators, thus eliminating any Joule heating or decoherence effects that arise from mobile electrons. We find that the energy spacings between the lowest few energy levels of the macroscopic quantum variable increase with the volume of the domain wall. Therefore, within an experimentally accessible range of material parameters, the energy gap between the qubit's logical states and the higher excited states can be as high as 10~K. For reasonable material parameters and operating temperatures of about 100~mK, we find that decoherence times associated with magnetic damping in the material are in the range of 10~ns, which are shorter than the state-of-the-art superconducting qubits~\cite{devoretSCI13}, and the number of Rabi flops within the coherence time scale is estimated to be $10^{2}$.  However, decoherence times in the $1~\mu$s range and $10^4$ Rabi flops within the coherence time can be achieved if the magnetic damping is in an ultra-low damping regime characterized by a Gilbert parameter $\al\sim10^{-7}$. This value is at least two orders of magnitude smaller than the currently reported values in magnetic insulators at room temperature.~\cite{heinrichPRL11,*hahnPRB13} However, a magnon-phonon theory for Gilbert damping~\cite{kasuyaPRL61} predicts that the Gilbert parameter vanishes linearly with temperature. Therefore, in the sub-Kelvin temperature regime relevant to the current proposal, it is possible that the damping parameter $\al$ is significantly smaller than the reported room temperature values. A recent experiment studied Gilbert damping over a wide temperature range and reported that disorder in the magnetic insulator can enhance $\al$ above the vanishing linear behavior at low temperatures.~\cite{flaigPRB17} The development of clean magnetic samples may thus be crucial in realizing the above ultra-low Gilbert damping regime.

The paper is organized as follows. In Sec.~\ref{sec2}, the domain wall quantum device is introduced, and a heuristic discussion of how the qubit can be realized in such a device is presented. In Secs.~\ref{sec2a} and \ref{cc}, a classical theory for the magnetic domain wall dynamics is derived in terms of the collective coordinates (i.e., the soft modes), and the magnetic damping is incorporated phenomenologically. In Sec.~\ref{tls}, the classical theory is quantized and the domain wall qubit is defined. Qubit control using ac magnetic fields is described in Sec.~\ref{sec4}, and the readout based on magnetic force microscopy is discussed in Sec.~\ref{sec5}. Various estimates for the physical properties of the qubit, including decoherence times and the upper temperature bound for quantum operation, are made in Sec.~\ref{sec6}. Conclusions are drawn in Sec.~\ref{sec7}.

\section{Model}
\label{sec2}
\begin{figure}[t]
\centering
\includegraphics[width=0.45\textwidth]{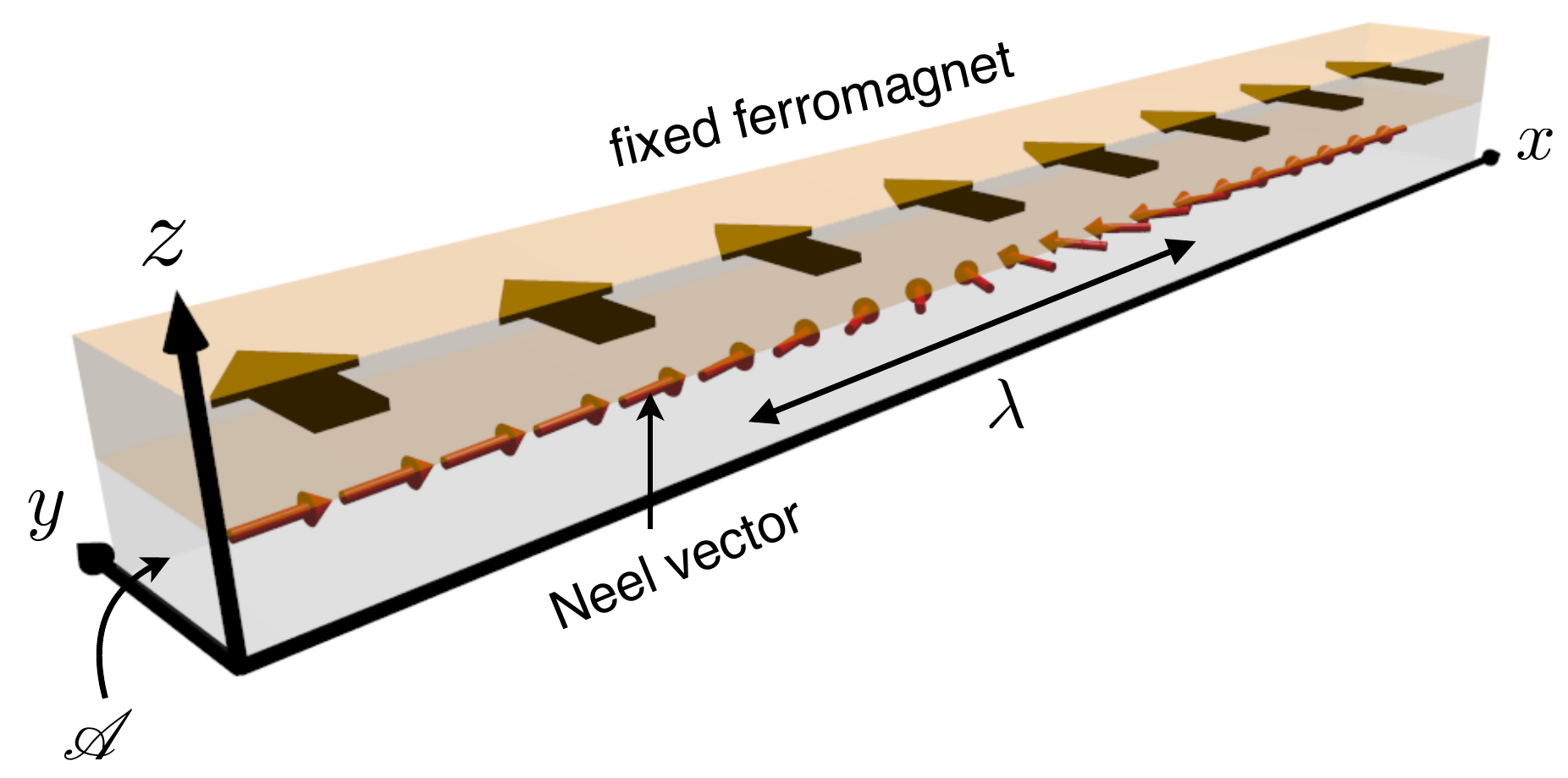}
\caption{The proposed macroscopic domain wall qubit. The antiferromagnet (the lower layer) contains a domain wall of width $\la$ and is exchanged-coupled to a mono-domain fixed ferromagnet. The cross-sectional area of the antiferromagnet is denoted by $\sA$.}
\label{fig1}
\end{figure}

The proposed macroscopic quantum device is shown in Fig.~\ref{fig1}. The insulating magnetic material (shaded in grey), hosting the domain wall (of width $\la$), is interfaced by another insulator with fixed single-domain ferromagnetic order (shaded in gold) with spins ordered in the positive $y$ direction. In this work, we consider the insulator hosting the domain wall to possess antiferromagnetic ordering, thus allowing us to ignore any stray field effects arising from the domain wall system and to streamline our discussion. We assume that the antiferromagnet possesses an uniaxial magnetic anisotropy $K$  along the $x$ axis such that the domain wall involves a gradual evolution of the N\'eel vector from the $+x$ to the $-x$ direction as shown and has an azimuth angle $\Phi$ (the angle the N\'eel vector makes relative to the positive $y$ axis) at the domain wall center. In this work, we assume an exchange coupling $J$ between the antiferromagnet and the ferromagnet that energetically favors a parallel alignment of the N\'eel vector and the ferromagnetic spins.~\cite{noguesJMMM99} 
Finally, we subject the antiferromagnet to a uniform magnetic field $\bB$ along the $y$ axis. Even in the presence of the exchange coupling and the external field, the N\'eel vector can, to a good approximation, maintain the domain wall spin structure illustrated in Fig.~\ref{fig1} if the uniaxial magnetic anisotropy (and the internal exchange scale of the antiferromagnet) is much greater than the exchange coupling and the Zeeman energy associated with the antiferromagnetic spins in the field.

Before delving into the detailed microscopic modeling of the system, let us first give a heuristic description of how the device in Fig.~\ref{fig1} can realize a macroscopic qubit. We first note that the domain wall can be described using the collective coordinate approach, in which the domain wall configuration and its dynamics are parametrized in terms of a few collective degrees of freedom representing the soft modes of the domain wall.~\cite{tretiakovPRL08} For a spatially pinned domain wall (which has been achieved for ferromagnetic domain walls by, e.g., introducing notches along the length of the ferromagnet or by modulating ferromagnet's material properties),~\cite{parkinSCI08} the domain wall's only relevant collective coordinate becomes $\Phi$. We then find that the (classical) Hamiltonian for the antiferromagnet in Fig.~\ref{fig1}, expressed in terms of this single coordinate $\Phi$, reduces to [c.f. Eq.~\eqref{ht}]
\beq
\label{hintro}
H_0=\frac{p_\Phi^2}{2M}+V(\Phi)\ ,
\eeq
which describes a ``particle" with effective mass $M$ (which depends on domain wall width $\la$ and various microscopic properties of the antiferromagnet) and coordinate $\Phi$ moving in a potential given by $V(\Phi)$. Here, $p_\Phi=M\dot\Phi$ is the momentum conjugate to $\Phi$ and is proportional to the angular velocity of the N\'eel vector at the center of the domain wall. The precise shape of the potential $V(\Phi)$ depends on the exchange coupling $J$, the magnetic field $B$, the domain wall width $\la$, and the cross-sectional area of the antiferromagnet $\sA$.

\begin{figure}[t]
\centering
\includegraphics[width=0.45\textwidth]{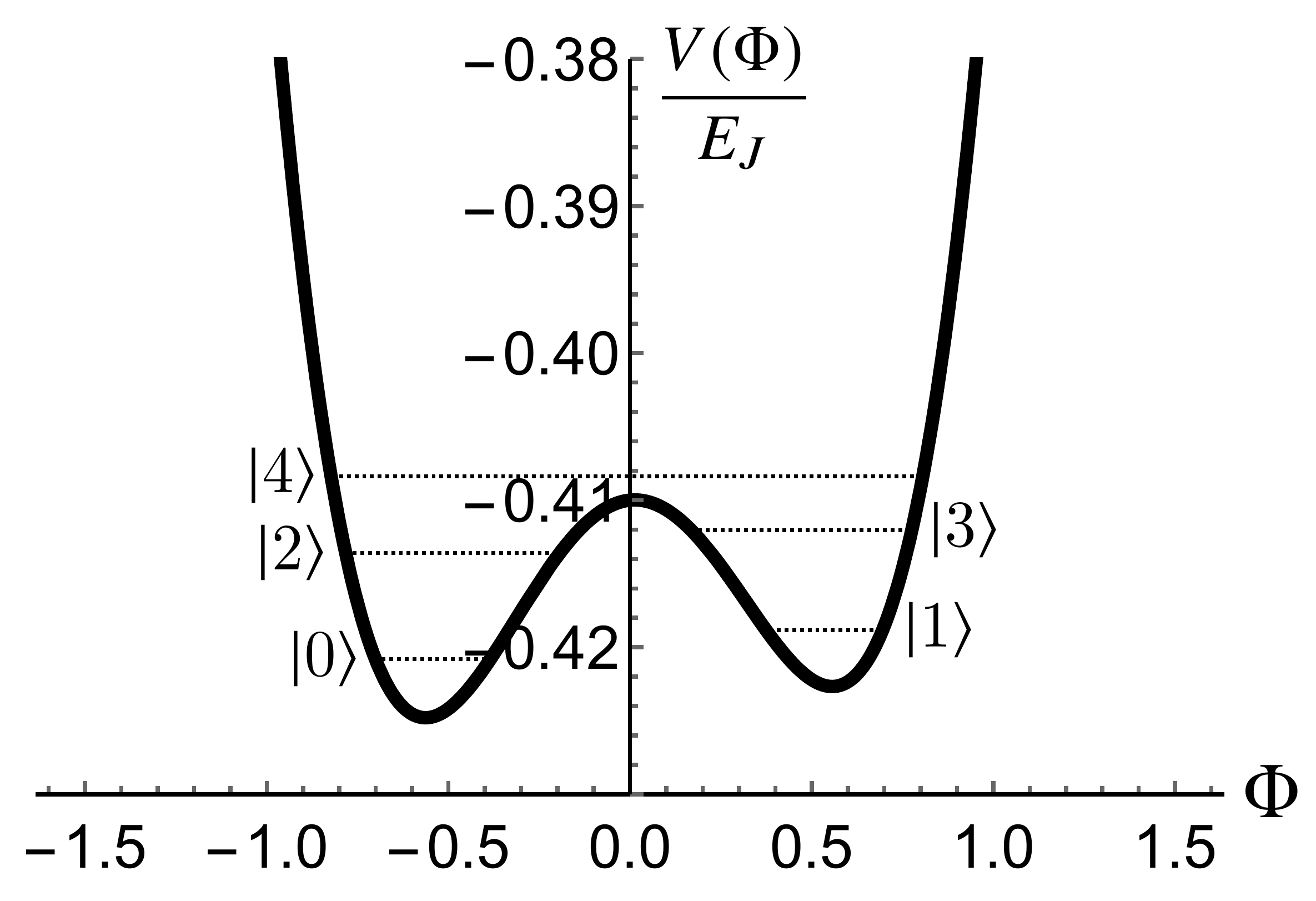}
\caption{Plot of potential $V(\Phi)$ as a function of $\Phi$. The five lowest quantum states have been numerically computed by solving Eq.~\eqref{h0q} for $\rho=1.18$ and $\be=0.002$.}
\label{fig2}
\end{figure}

For reasonable magnitudes of these various physical parameters, the potential $V(\Phi)$ has a double-well profile as a function of $\Phi$ as shown in Fig.~\ref{fig2}, thus providing the two minima that define the two-dimensional logical space of the qubit. The qubit is eventually defined by promoting the conjugate variables $(\Phi,p_\Phi)$ to quantum operators, i.e., $p_\Phi\rightarrow-i\hbar\pd_\Phi$, and quantizing Eq.~\eqref{hintro}. For a set of parameters (defined later on), the five lowest quantum states are shown as dotted lines in Fig.~\ref{fig2}. The qubit states are defined by the two lowest states, corresponding to two stable azimuth orientations of the N\'eel vector at the center of the domain wall. In this work, we identify a regime in which the thermal fluctuations are reduced well below the level of quantum fluctuations and the damping is low enough so that the lifetime of the quantum energy levels is much greater than the Rabi time for coherent oscillations. In this (so-called quantum) regime, the macroscopic quantum domain wall variable $\Phi(t)$ is expected to exhibit all the quantum effects predicted by elementary quantum mechanics. In later sections, we discuss in detail how arbitrary single-qubit rotations can be achieved by applying small ac magnetic field pulses. We now provide the theoretical underpinnings for the above heuristic discussion. 

\subsection{Theory}
\label{sec2a}
We first begin by developing a model for the insulator hosting a single domain wall. For a bipartite antiferromagnet, an effective long-wavelength theory can be developed in terms of two continuum variables $\bn$ and $\bm$, which respectively denote the staggered and uniform components of the spin density.~\cite{andreevSPU80} We consider a quasi-one-dimensional antiferromagnet, whose axis lies collinear with the $x$ axis and whose cross-sectional area is $\sA$ (we assume homogeneity of $\bn$ and $\bm$ across the cross-section). Its effective Lagrangian, obeying sublattice exchange symmetry (i.e., $\bn\rightarrow-\bn$ and $\bm\rightarrow\bm$), then may be written as~\cite{andreevSPU80}
\beq
\label{l}
L=\sA s\int dx\ \bm\cdot(\dot\bn\times\bn)-U[\bm,\bn]\ ,
\eeq
where $s$ denotes the spin density on one of the sublattices. Here, the continuum variables satisfy the constraints $|\bn|=1$ and $\bn\cdot\bm=0$, and strong local N\'eel order implies $|\bm|\ll1$. The potential energy $U[\bm,\bn]$, to gaussian order in $\bm$, may be expressed as 
\beq
U[\bm,\bn]=\sA\int dx\ \round{\frac{\bm^2}{2\x}+\bb\cdot\bm}+U[\bn]\ ,
\eeq
where $\x$ denotes the spin susceptibility for a uniform external magnetic field $\bB$ applied transverse to the axis of the N\'eel  order, and $\bb=\g s\bB$ (with $\g$ the gyromagnetic ratio). Throughout this work, the field $\bB$ will be confined within the $yz$ plane. The uniform component $\bm$ may be eliminated from the Lagrangian if we note it is merely a slave variable defined by the staggered variable, i.e., $\bm=\x s\dot\bn\times\bn-\x\bn\times(\bb\times\bn)$. Upon eliminating $\bm$, the effective Lagrangian~\eqref{l} reduces to
\beq
\label{l2}
L=\frac{\sA\x}{2}\int dx\ \round{s\dot\bn+\bn\times\bb}^2-U[\bn]\ .
\eeq
Within the Lagrangian formalism, viscous spin losses in the antiferromagnet can be represented via the Raleigh dissipation function $R$, which for the antiferromagnet may be given by
\beq
\label{gd}
R=\al\frac{\sA s}{2}\int dx\ |\dot\bn|^2\ ,
\eeq
where $\al$ is the dimensionless Gilbert damping constant. For $\hbar/\tau,ba^3\ll E_{\rm ex}$, where $\tau$ is the time scale for the N\'eel dynamics, $E_{\rm ex}$ is the internal antiferromagnetic exchange scale and $a$ is the lattice scale, the contribution to the Raleigh function from the uniform component may be neglected.

Fluctuations of the N\'eel variable at finite temperature can be modeled by introducing a stochastic field $\boldsymbol{\eta}$. Through the fluctuation-dissipation theorem, the correlator for the cartesian components of this Langevin field relates to the Gilbert damping constant $\al$ as~\cite{landauBOOKv5}
\beq
\langle \eta_i(x,\w)\eta_j(x',\w')\rangle=\frac{2\p\al\sA s\hbar\w}{\tanh(\hbar\w/2k_BT)}\de_{ij}\de(x-x')\de(\w+\w')\ .
\eeq
Defining the momentum $\bpi$ conjugate to $\bn$ via Eq.~\eqref{l2}, 
\beq
\bpi\equiv\frac{\de L}{\de \dot\bn}=\sA\x s\round{s\dot\bn+\bn\times\bb}\ ,
\eeq
the Hamilton's equation corresponding to its dynamics [including damping Eq.~\eqref{gd} and the stochastic contribution] reads
\beq
\label{he}
\dot\bpi=-\frac{\de U[\bn]}{\de\bn}-\frac{\de R}{\de\dot\bn}+\boldsymbol{\eta}\ .
\eeq
For the potential for the staggered variable, we take
\beq
\label{un}
U[\bn]=\sA\int dx\ \square{\frac{A}{2}(\pd_x\bn)^2-\frac{K}{2}n_x^2-Jn_y}\ ,
\eeq
where $A$ is the stiffness constant associated with the N\'eel vector, $K>0$ is the easy-axis anisotropy along the wire axis, and $J$ is an exchange coupling between the local N\'eel vector and a hard mono-domain ferromagnet (see Fig.~\ref{fig1}) whose spin moments are assumed to be fixed in the $y$ direction.

\subsection{Domain wall dynamics}
\label{cc}
The dynamics for the antiferromagnetic domain wall can be derived by following the standard collective coordinate approach.~\cite{tvetenPRL13,kimPRB15,tvetenPRB16,rodriguesPRB17} The static domain wall solution can be obtained by extremizing Lagrangian~\eqref{l2} in the static limit. In the limit of strong easy-axis anisotropy $K\gg b^2\x,J$, we may well-approximate the domain wall solution with the solution at $\bB=0$ and $J=0$, i.e.,
\beq
\label{n0}
\bn=\{\cos\thi(x),\sin\thi(x)\cos\Phi,\sin\thi(x)\sin\Phi\}\ ,
\eeq
where $\cos\thi(x)=\tanh[(x-X)/\la]$, $\la\equiv\sqrt{A/K}$ quantifies the width of the wall, and the two ``soft modes" $X$ and $\Phi$ respectively denote the position of the domain wall and its azimuth angle within the $yz$ plane. These variables describe the dynamics of the domain wall well if the dynamics is slower than the spin-wave gap set by the easy-axis anisotropy.~\cite{tretiakovPRL08} A strong domain wall pinning (at, e.g., $X=0$) effectively gaps out the positional soft mode $X$ such that for dynamics slow compared to this gap, $\Phi$ remains as the only relevant mode. The damped Langevin dynamics can be obtained by taking the inner product of both sides of Eq.~\eqref{he} by $\pd\bn/\pd\Phi$, inserting Eq.~\eqref{n0} into the equation and integrating over $x$,
\beq
\label{phid}
M\ddot\Phi+\frac{M}{\tau}\dot\Phi=-\frac{\pd V(\Phi)}{\pd \Phi}+f(t)\ ,
\eeq
where the stochastic force $f(t)$ obeys the correlator
\beq
\label{fdt}
\langle f(t)f(0)\rangle=\int\frac{d\w}{2\p}S_f(\w)e^{-i\w t}\ ,
\eeq
where $S_f(\w)=(M\hbar\w/\tau)\coth(\hbar\w/2k_BT)$. The dynamics for $\Phi$, i.e., Eq.~\eqref{phid}, is equivalent to the dynamics of a particle with mass $M\equiv 2\la\sA\x s^2$ and a damping constant proportional to the inverse relaxation time $\tau\equiv\x s/\al$, moving in a potential given by
\beq
\label{vphi}
V(\Phi)=-E_J\square{\cos\Phi-\frac{\rho}{2}(\cos\Phi+\be\sin\Phi)^2}\ ,
\eeq
where $E_J\equiv\p\la\sA J$ is the exchange energy stored within the domain wall volume, $\rho\equiv2\x b_y^2/\p J$ parametrizes the strength of the magnetic field along the $y$ axis, and $\be\equiv B_z/B_y$ quantifies the canting of the field away from the axis. 

\section{Domain Wall Qubit}
\label{tls}
Noting that the momentum conjugate to $\Phi$ is $p_\Phi\equiv M\dot\Phi$, the Hamiltonian corresponding to the dynamics Eq.~\eqref{phid} (ignoring the damping but including the stochastic contribution) is given by
\beq
\label{ht}
H(t)=\frac{p_\Phi^2}{2M}+V(\Phi)-\Phi f(t)\equiv H_0-\Phi f(t)\ .
\eeq
We only consider the regime of $\rho\gtrsim1$ and small field canting in the $z$ direction, i.e., $0\le\be\ll1$. The potential $V(\Phi)$ is plotted as a function of $\Phi$ in Fig.~\ref{fig2} for, e.g., $\rho=1.18$ and $\be=0.002$. At $\be=0$, the potential has two degenerate minima separated by a central barrier, which disappears as $\rho\rightarrow1$. A finite $\be$ removes this degeneracy and raises the right minimum to a higher energy than the left minimum for $\be>0$ (and vice versa for $\be<0$). The two minima physically correspond to two stable azimuth orientations of the N\'eel vector at the center of the domain wall.

The deterministic part of the Hamiltonian $H_0$ can be quantized by promoting the conjugate variables to operators, i.e., $p_\Phi\rightarrow-i\hbar\pd_\Phi$,
\beq
\label{h0q}
\hH_0=-E_M\pd_\Phi^2+V(\Phi)\ ,
\eeq
where $E_M\equiv\hbar^2/2M$. For the system to operate in the so-called {\em phase} regime, we require
\beq
\frac{E_J}{E_M}=4\p S^2\frac{B_J}{B_{\rm ex}}\gg1,
\eeq
where the prefactor $S\equiv\la\sA s/\hbar$ gives the total spin (in units of $\hbar$) on one sublattice located inside the domain wall volume $\la\sA$, $B_J$ is the exchange field corresponding to the coupling between the ferromagnet and the antiferromagnet, and $B_{\rm ex}$ is the internal antiferromagnet exchange field. While typically $B_J\ll B_{\rm ex}$, a large enough total spin inside the domain wall volume would satisfy the above inequality.  

At low temperatures, the quantization of energy levels near the two potential minima becomes important as shown in Fig.~\ref{fig2}. By solving the Schr\"odinger equation~\eqref{h0q} numerically, the lowest five states are shown by dashed lines in Fig.~\ref{fig2}. If the system is in the lowest energy state of a well, spontaneous thermal excitation up into higher energy levels can be avoided by lowering the temperature to $T\ll(E_2-E_0)/k_B\sim(E_3-E_1)/k_B\equiv T_{\rm max}$, where $E_n$ are the energy eigenstates for state $\ket{n}$. We find from Fig.~\ref{fig2} that $T_{\rm max}\sim0.01E_J/k_B$.

Considering a finite but small $\be$ (to linear order in $\be$), the splitting between the two lowest energy levels reads $\hbar\w_{10}\equiv E_1-E_0\approx2E_J\be\sqrt{1-\rho^{-2}}$. We assume $\hbar\w_{10}/k_BT_{\rm max}\ll1$ so that inter-well resonances are avoided.

\section{Coherent control of domain wall qubit}
\label{sec4}
To discuss how the qubit can be controlled, we introduce a small oscillating component to the magnetic field in the $y$ direction, i.e., $b_y(t)=b_y[1+\z(t)]$, where $b_y$ is the original dc component already accounted for above, and (dimensionless) $\z(t)$ quantifies the small oscillating component. From Eq.~\eqref{vphi}, to lowest order in small quantities $\z(t),\be\ll1$, this oscillating field gives rise to a new contribution, 
\beq
\label{hht}
h(t)=E_J\rho\cos^2\Phi\z(t)\ ,
\eeq
to the (deterministic) classical Hamiltonian $H_0$ in Eq.~\eqref{ht}.

\begin{figure}[t]
\centering
\includegraphics*[scale=0.3]{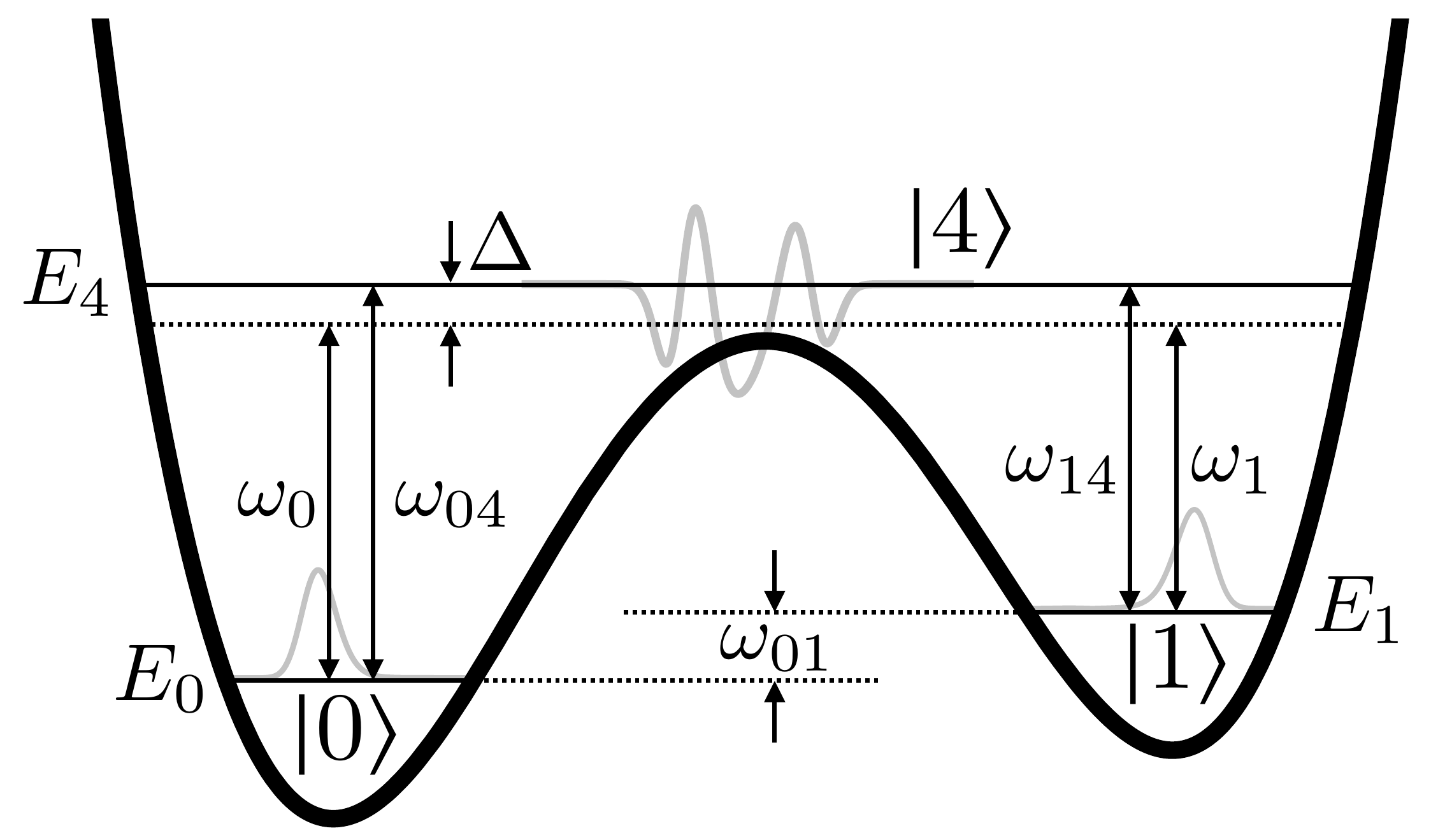}
\caption{Three quantum levels (the two logical states $\ket{0}$ and $\ket{1}$ and an auxiliary excited state $\ket{4}$) involved in qubit control. Drive frequencies, $\w_0$ and $\w_1$, are symmetrically detuned from energy separations, $\w_{04}$ and $\w_{14}$, to the excited state with detuning $\w_{04}-\w_0=\w_{14}-\w_1\equiv\De/\hbar$. We assume $\w_0$ and $\w_1$ are far-detuned from all the other levels so that we may consider this three-level subsystem.}
\label{fig3}
\end{figure}

For qubit control, we consider an oscillating component $\z(t)=2[\z_0\cos(\w_0 t)+\z_1\cos(\w_1 t)]$, where frequencies $\w_0$ and $\w_1$ are detuned from energy separations $\w_{04}$ and $\w_{14}$ to an excited state $\ket{4}$ with a (symmetric) detuning $\w_{04}-\w_0=\w_{14}-\w_1\equiv\De/\hbar$. We assume $\w_0$ and $\w_1$ are far-detuned from all the other levels so that we may consider a three-level subsystem as shown in Fig.~\ref{fig3}. Here, the use of state $\ket{4}$ is somewhat arbitrary; in principle, any excited state can be used as long as there are appreciable transition matrix elements [generated by Eq.~\eqref{hht}] between the state and the logical states and the resonance condition above can be satisfied. We further assume that $\w_0$ is far-off resonant with $\w_{01}$ and $\w_{14}$ (i.e., $|\w_{01}-\w_0|\gg\W^{(0)}_{01}$ and $|\w_{14}-\w_0|\gg\W^{(0)}_{14}$, where $\W^{(0)}_{01}$ and $\W^{(0)}_{14}$ are Rabi frequencies for on-resonance $0\leftrightarrow1$ and $1\leftrightarrow4$ transitions, respectively, generated by the $\z_0$ pulse) and that $\w_1$ is far-off resonant with $\w_{01}$ and $\w_{04}$ (i.e., $|\w_{01}-\w_1|\gg\W^{(1)}_{01}$ and $|\w_{04}-\w_1|\gg\W^{(1)}_{04}$, where $\W^{(1)}_{01}$ and $\W^{(1)}_{04}$ are Rabi frequencies for on-resonance $0\leftrightarrow1$ and $0\leftrightarrow4$ transitions, respectively, generated by the $\z_1$ pulse). Invoking the rotating wave approximation, the effective three-level Hamiltonian can then be written as
\beq
\hH_{\rm eff}^{(3)}=\sum_{n=0,1,4}E_n\ket{n}\bra{n}+\sum_{n=0,1}\hbar\W_{n}e^{i\w_nt}\ket{n}\bra{4}+h.c.\ ,
\eeq
where $\W_{n}=E_J\rho\z_n|\bra{n}\cos^2\Phi\ket{4}|/\hbar$. For large detuning, i.e., $\De\gg\W_0,\W_1$ (and if the initial state is in the logical subspace), the excited state is expected to play a small role in the dynamics of the system because they are far-off-resonantly and weakly coupled to the logical subspace. In this case, we may eliminate the (irrelevant) excited state in the adiabatic approximation, which can be carried out by first going to a rotating frame defined by $\hU=e^{-i\hA t/\hbar}$ and $\hA=\diag\{E_0,E_1,E_4-\De\}$ and solving the Schr\"odinger equation in the new frame under the constraint $\pd_t\ket{4}\approx0$. After elimination, the Schr\"odinger equation for the qubit subspace reads $i\hbar\pd_t\Psi(t)=\hH^{(2)}_{\rm eff}\Psi(t)$, where the effective two-level Hamiltonian is given by
\beq
\hH_{\rm eff}^{(2)}=-\frac{\hbar}{2}\W_z\hsig_z-g\hsig_x\ ,
\eeq
with $\W_z=\hbar(\W_0^2-\W_1^2)/\De$ and $g=\hbar^2\W_0\W_1/\De$. Arbitrary single-qubit rotations can be achieved by controlling the pulse amplitudes $\z_0$ and $\z_1$.

\section{Domain wall qubit readout}
\label{sec5}
The two states of the qubit correspond to two classically distinct orientations of the N\'eel vector within the $yz$ plane at the center of the domain wall. Since the two minima of the double-well potential in Fig.~\ref{fig2} are located at $\Phi=\Phi_0\equiv\cos^{-1}(1/\rho)$, the $\ket{0}$ state corresponds to $\bn=(0,\cos\Phi_0,-\sin\Phi_0)$ and the $\ket{1}$ to $\bn=(0,\cos\Phi_0,\sin\Phi_0)$. At the end of the computation, the N\'eel vector should be in one of the two classical configurations and should remain in the same configuration even when the field component in the $z$ direction is adiabatically switched off. The presence of the remaining magnetic field in the $y$ direction implies that a small total spin component $\bm$ is present in the antiferromagnet since $\bm=-\x\bn\times(\bb\times\bn)$. In particular, we expect a magnetization of approximately $\bM_s=\x\g s b_y\bn\times(\ey\times\bn)$ within the domain wall volume. If the qubit state is in $\ket{0}$, $\bM_s=M_s(0,\sin\Phi_0,\cos\Phi_0)$ with $M_s=\x\g sb_y\sin\Phi_0$, and $\bM_s=M_s(0,\sin\Phi_0,-\cos\Phi_0)$ in state $\ket{1}$.

\begin{figure}[t]
\centering
\includegraphics*[scale=0.2]{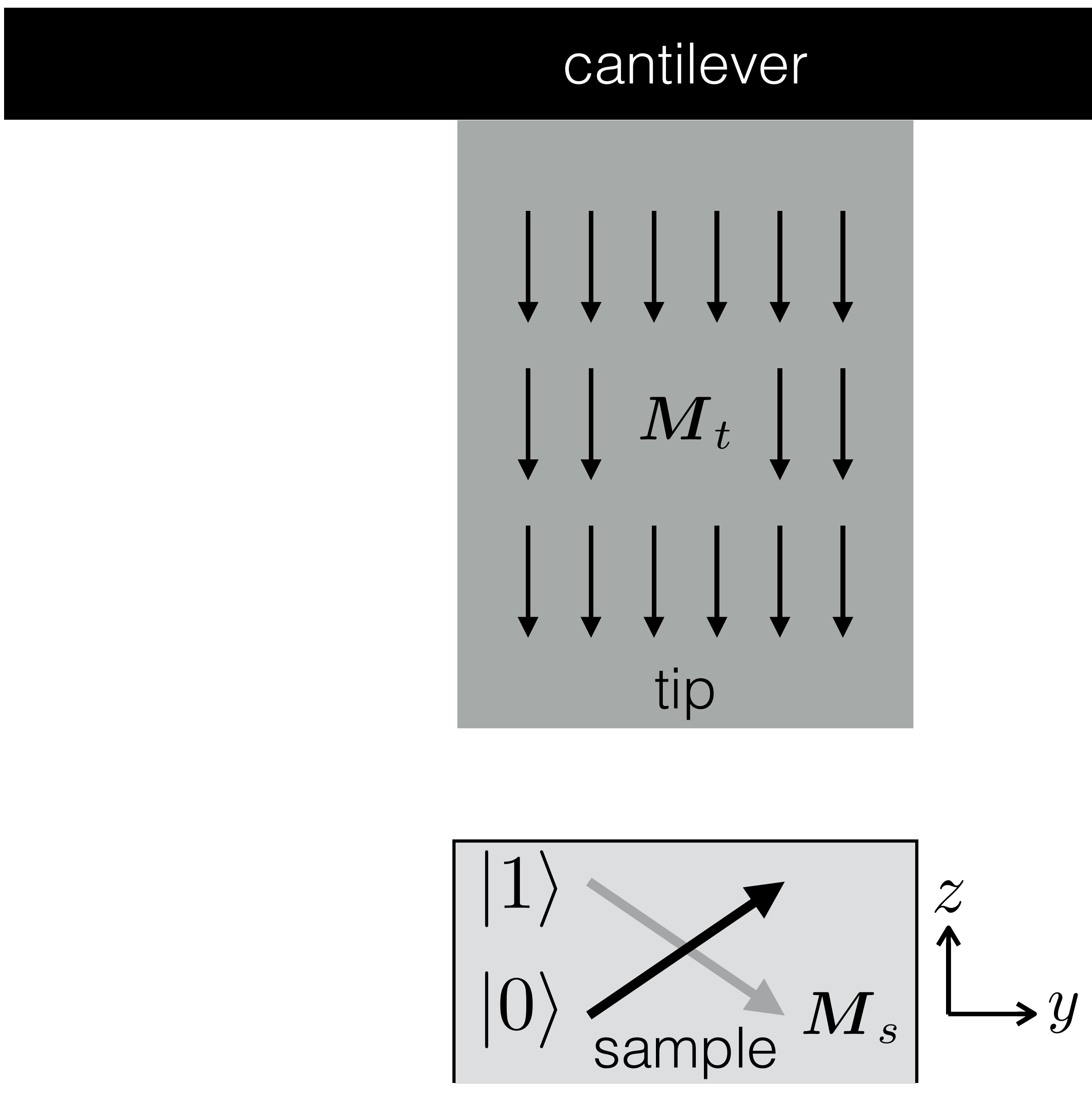}
\caption{A cartoon showing how magnetic force microscopy can be used to probe the magnetic state of our qubit.}
\label{fig4}
\end{figure}

We now estimate the sensitivity required to probe the orientation of $\bM_s$ using, e.g., magnetic force microscopy (MFM).~\cite{porthunJMMM98} An MFM consists of a magnetized tip placed at the end of a cantilever as shown in Fig.~\ref{fig4}. In the dynamic mode, the cantilever (assumed here to be oriented parallel to the $xy$ plane) is oscillated in the $z$ direction at its resonance frequency $\W_{\rm res}=\sqrt{k_{\rm eff}/m_t}$, where $m_t$ is the mass of the tip and the cantilever combined. The effective spring constant $k_{\rm eff}=k_0-\pd F_z/\pd z$, where $k_0$ is the cantilever spring constant and $F_z$ is the $z$ component of the force acting on the magnetic tip (with magnetization $\bM_t$) when it is brought in close proximity to the magnetic sample (with magnetization $\bM_s$).~\cite{porthunJMMM98} The force $\bF$ acting on the magnetic tip is given by $\bF=\mu_0\int dV_t\ \boldsymbol{\nabla}(\bM_t\cdot\bH_s)$, where the volume integral is performed over the tip and $\bH_s$ is the device's stray field acting on the tip. If we consider a magnetic tip with uniform magnetization $\bM_t=M_t\ez$, the ($z$ component of the) force simplifies to $F_z=\mu_0M_t\int dV_t\ \pd_zH^z_s$. In order to estimate $F_z$, we model our device as a rectangular prism with a uniform magnetization $\bM_s$ and occupying the region $-\la/2\le x\le \la/2$, $-w/2\le y\le w/2$ ($w$ being the width of the antiferromagnet, which we assume to obey $w^2=\sA$) and $-w\le z\le 0$, and model our tip as a rectangular prism occupying the region $-w_t/2\le x\le w_t/2$, $-w_t/2\le y\le w_t/2$ and $z_0\le z\le z_0+l_t$. Noting that $\bH_s(\bx)=-\boldsymbol{\nabla}\f_m(\bx)$, where the magnetostatic scalar potential here is given by~\cite{hartmannARMS99}
\beq
\f_m(\bx)=\frac{1}{4\p}\int\frac{d^2\bs'\cdot\bM_s(\bx')}{|\bx-\bx'|}\ ,
\eeq
and $\bs'$ is the outward normal vector from the sample surface, the relative shift in the resonance frequency corresponding to the two states then becomes
\beq
\label{rfs}
\frac{\De\W_{\rm res}}{\sqrt{k_0/m_t}}=\frac{\mu_0M_t}{2k_0}\int dV_t\ \pd^3_z\f_m(\bx)\ .
\eeq

\section{Decoherence model}
\label{dec}
Decoherence can arise due to viscous spin losses [recall Gilbert damping incorporated via Eq.~\eqref{gd}], which, from the fluctuation-dissipation theorem, gave rise to the stochastic contribution in the classical Hamiltonian~\eqref{ht} for the collective coordinates. Additional decoherence can arise from fluctuations in the static background magnetic field $b_y$ [which we denote $\de b_y(t)$] and in the amplitudes of the pulse fields $\z_{0,1}$ [which we denote $\de\z_{0,1}(t)$]. We focus here on the noise due to Gilbert damping, as estimating decoherence rates due to field fluctuations requires specifying their spectrum, which depends on the precise details of how the fields are generated in our device and such details are beyond the scope of the current work. 

By numerically solving Eq.~\eqref{h0q} for $\rho=1.18$ and $\be=0.002$ (the parameters used in Sec.~\ref{tls}), we find that the off-diagonal elements of $|\bra{i}\Phi\ket{j}|$ [see Eq.~\eqref{ht}] are small compared to the diagonal terms so we ignore them. Upon performing the adiabatic elimination and linearizing in the fluctuations, we may write the stochastic contribution to the effective two-level Hamiltonian as  
\beq
\label{dw}
\de\hH_{\rm eff}^{(2)}(t)\approx\Phi_0f(t)\hsig_z\ .
\eeq

To study decoherence, we go to a new eigenbasis, defined by a total effective field with $z$ component $-\hbar\W_z/2$ and $x$ component $-g$, which can be achieved by a unitary transformation $\tilde\hU=e^{-i(\p/2-\thi)\hsig_y/2}$ with $\thi=\tan^{-1}(\hbar\W_z/2g)$. The effective two-level Hamiltonian in the new basis then reads
\beq
\tilde\hH^{(2)}_{\rm eff}=-\frac{\hbar}{2}\tilde\W_z\hsig_z+\Phi_0f(t)\round{\sin\thi\hsig_z-\cos\thi\hsig_x}\ ,
\eeq
where $\hbar\tilde\W_z=2[(\hbar\W_z/2)^2+g^2]^{1/2}$. Then using standard results for noise and decoherence in quantum two-level systems, we find for the longitudinal relaxation and dephasing rates,~\cite{ithierPRB05}
\beq
\label{g1g2}
\G_1=\frac{2\Phi_0^2\cos^2\thi}{\hbar^2}S_f(\tilde\W_z)\ ,\quad\G_2=\frac{\G_1}{2}+\G_\varphi\ ,
\eeq
where the pure dephasing rate is
\beq
\label{gphi}
\G_\varphi=\frac{2\Phi_0^2\sin^2\thi}{\hbar^2} S_f(0)\ .
\eeq

\section{Discussion}
\label{sec6}
To make numerical estimates, we use a domain wall width of $\la=10$~nm, and consider an antiferromagnet with a simple cubic lattice of spin-1's, lattice constant $a=5\AA$, width $w=10$~nm and $\x=a^3/\hbar\g B_{\rm ex}$ with $B_{\rm ex}=90$~T. Assuming exchange field between the antiferromagnet and the ferromagnet of $B_J=50$~mT [and noting that $J=(\hbar\g B_J/a^3)$ and $b_y=\hbar\g B_y/a^3$], ensuring $\rho=1.18$ requires $B_y\approx2$~T, which is relatively large but still experimentally manageable. This also implies that the uniaxial anisotropy for the N\'eel vector must obey $K\gg7000$~J/m$^3$ in order for the device to remain in the strong anisotropy regime (i.e., $K\gg \x b_y^2,J$). Since $E_J/k_B=\p\la\sA J/k_B\approx2000$~K, spontaneous thermal excitations to higher energy levels (recall discussion in Sec.~\ref{tls}) are suppressed for temperatures $T\ll T_{\rm max}=0.01E_J/k_B\approx 20$~K. For $\be=0.002$ (used in the numerics in Fig.~\ref{fig2}), the inter-well resonances are avoided since $2E_J\be\sqrt{1-\rho^{-2}}\approx4~\mbox{K}\ll T_{\rm max}$. Finally, the device is ensured to be within the phase regime since $E_J/E_M\sim10^5$ for these parameters.

The frequencies of the ac magnetic field pulses $\w_0$ and $\w_1$ needed for qubit control are set by energy level spacings $E_4-E_0$ and $E_4-E_1$, respectively (see Fig.~\ref{fig3}), which are both of order $k_BT_{\rm max}$. For $T_{\rm max}=20$~K (as above), we obtain $\w_0/2\p\sim\w_1/2\p\approx 350$~GHz.

The relative resonance frequency shift that is necessary for qubit readout was given by Eq.~\eqref{rfs}. For a magnetic tip with dimensions 100~nm~$\times$~100~nm~$\times$~500~nm placed directly above the sample at a height of 5~nm, $M_t=1000$~emu/cm$^3$ and $k_0\sim0.1$~N/m (and noting that here $M_s^z\approx1600$~A/m),~\cite{hartmannARMS99} we require relative frequency shift of $|\De\W_{\rm res}/\sqrt{k_0/m_t}|\sim10^{-5}$ in order to detect the qubit state. For cantilevers with resonance frequencies in the range of 100~kHz, the frequency resolution necessary is in the 1~Hz range.

The noise spectrum for the stochastic field $f(t)$ was specified by fluctuation-dissipation theorem in Eq.~\eqref{fdt}. From Eqs.~\eqref{g1g2} and \eqref{gphi}, we then obtain
\begin{equation}
\G_1=\frac{2\Phi_0^2S_f(\tilde\W_z)}{\hbar^2\square{1+\round{\frac{\hbar\W_z}{2g}}^2}},\ \ 
\G_2=\frac{\Phi_0^2}{\hbar^2}\square{\frac{S_f(\tilde\W_z)}{1+\round{\frac{\hbar\W_z}{2g}}^2}+\frac{2S_f(0)}{1+\round{\frac{2g}{\hbar\W_z}}^2}}\ .
\end{equation}
For concreteness, we compute the decoherence rates for $\hbar\W_z=2g=\hbar^2\W_0\W_1/\De$. Since $\De\sim0.01E_J$ and the large detuning limit implies $\hbar\W_n\ll\De$, we may take $g\sim10^{-4}E_J$. Since $k_BT\ll0.01 E_J$ as well (for quantum operation), we assume $k_BT\sim g$ (which implies $T\sim 100$~mK), so that
\beq
\label{G1}
\G_1\sim\G_2\sim\frac{\al\W_R\la\sA s}{\hbar}
\eeq
(up to constants of order 1), where $\W_R=g/\hbar$ is the qubit Rabi frequency. Equation~\eqref{G1} shows that while increasing the domain wall volume $\la\sA$ (i.e., making the qubit more macroscopic) is beneficial in that it increases the upper temperature bound $T_{\rm max}$ for quantum operation (since $T_{\rm max}\propto E_J\propto\la\sA$), it is unfavorable in that it decreases the coherence time of the qubit. Using a Gilbert damping parameter of $\al\sim10^{-5}$, and since $\la\sA s/\hbar\sim10^3$ and $g\sim 10^{-4}E_J$, we find a decoherence time of $\tau_{\rm dec}\sim10$~ns. The number of Rabi flops within the coherence time scale can be estimated by $\hbar/\al\la\sA s$. With the above parameters, we have $\hbar/\al\la\sA s\sim10^{2}$ which is within the range of $10^{2}$ to $10^{4}$ that in principle could allow for quantum error correction procedures. 

In order to achieve longer decoherence times and a lower threshold of $10^4$ Rabi flops within the coherence time scale, one may reduce $\al$ and/or $\la\sA$, i.e., the domain wall volume. Reducing the domain wall volume leads to the reduction in $E_J$ and hence the operational temperature. This may not be desirable as this reduces the operational temperature further below the 100~mK range. The desired benchmark can also be reached if Gilbert damping is decreased to the ultra low damping regime of $\al\sim10^{-7}$, which gives decoherence times in the $1~\mu$s range and $10^4$ Rabi flops within the coherence time scale. The value of $\al\sim10^{-7}$ is much smaller than the currently reported values in magnetic insulators at room temperature (e.g., $\al\sim10^{-4}-10^{-5}$ for yttrium iron garnet.~\cite{heinrichPRL11,*hahnPRB13}) However, a magnon-phonon theory for Gilbert damping by Kasuya and LeCraw~\cite{kasuyaPRL61} predicts that Gilbert damping vanishes linearly with temperature. Therefore, in the sub-Kelvin temperature regime relevant to the current proposal, it is possible that the damping parameter $\al$ is significantly smaller than the reported room temperature values. A recent experiment studied Gilbert damping from 5~K to 300~K, and reported that $\al$ vanishes linearly with temperature for $T>100$~K but develops a peak below $100$~K.~\cite{flaigPRB17} The experiment attributes the dominant relaxation mechanism at low temperatures to impurities, suggesting that the use of clean magnetic samples may be crucial in realizing the above ultra-low Gilbert damping regime. 

\section{Conclusion and Future Directions}
\label{sec7}
This work introduces a way to coherently control and readout a non-trivial macroscopic quantum state stored in a topological defect of a magnetic insulator. We find that the logical energy states of the qubit are separated from all excited states by an energy gap that is proportional to the volume of the domain wall $\la\sA$. This, in principle, allows us to exploit the high magnetic ordering temperature and increase the temperature window for quantum operation by going to a larger domain wall volume. We also find, however, that the qubit decoherence rate is also proportional to the domain wall volume, so the quantum coherence in the device would be compromised in large systems. For reasonable domain wall volumes and temperatures in the $100$~mK range, we find that the decoherence time and the number of Rabi flops within the coherence time can, respectively, reach 1~$\mu$s and $10^4$ if Gilbert damping is reduced into the range of $\al\sim10^{-7}$. This ultra-low Gilbert damping may be attainable within the above temperature range and for very clean magnetic samples.

It is interesting to consider how the final state can be read out using means other than MFM. Recently, a scanning probe microscopy technique was developed using a single-crystalline diamond nanopillar containing a single nitrogen-vacancy (NV) color center.~\cite{maletinskyNATN12} By recording the magnetic field along the NV axis produced by the magnetization pattern in the film, all three components of the magnetic field were reconstructed.~\cite{dovzhenkoCM16} This technique allows one to obtain the underlying spin texture from a map of the magnetic field,~\cite{dovzhenkoCM16} so it may serve as an alternative way to determine the direction of the magnetization at the domain wall center.

In principle, two coupled qubits can be realized in the above proposal by placing an additional domain wall inside the same antiferromagnet in Fig.~\ref{fig1} and bringing the two domain walls spatially close together. While the precise quantitative theory for the domain wall coupling in an antiferromagnet is beyond the scope of the current work, we can heuristically see how a coupling between the two macroscopic qubit variables, i.e., the angles $\Phi_1$ and $\Phi_2$ (respectively corresponding to the azimuth angles at the centers of the domain walls for qubits 1 and 2) can be engendered. N\'eel order stiffness of the antiferromagnet should lead to a smaller interaction energy for two nearby domain walls with $\Phi_1=\Phi_2$ than for $\Phi_1\ne\Phi_2$, since in the latter case the N\'eel texture must twist from $\Phi_1$ to $\Phi_2$ as one moves from qubit 1 to qubit 2. This should give rise to an interaction energy between two domain walls that depends on the two angles, i.e., $E_{\rm int}=E_{\rm int}(\Phi_1,\Phi_2)$. By projecting the resulting interaction energy to the qubit logical space, a qubit coupling should then arise. While the interaction energy between two transverse domain walls has been investigated theoretically in a ferromagnetic wire,~\cite{kruegerJPCM12} a similar calculation for antiferromagnets has not been done to the best of our knowledge. It is indeed interesting to study domain wall coupling in antiferromagnetic systems and how the result engenders a two-qubit coupling in the current proposal.

Finally, while an antiferromagnetic domain wall qubit is more attractive for it reduces the effects of stray fields (which may generate unwanted crosstalk between qubits when multiple devices are coupled), we note here that an analogous domain wall qubit can be realized using a ferromagnetic insulator. To this end, one can consider a ferromagnetic domain wall as in Fig.~\ref{fig1} with a parabolic pinning potential with curvature $\ka$ pinning its center at, e.g., $X=0$; a combination of external fields and an appropriate magnetic anisotropy within the $yz$ plane can be used to construct a double-well potential as a function the domain wall's azimuth angle $\Phi$ (the angle of the spin within the $yz$ plane at the center of the domain wall) as shown in Fig.~\ref{fig2}. If the dynamics is projected down to the two soft modes of the domain wall, i.e., domain wall position $X$ and the angle $\Phi$, the Hamiltonian exactly analogous to Eq.~\eqref{hintro} can be constructed with the conjugate momentum $P_\Phi$ given by $X$ and with the effective mass $M$ given by the (inverse of the) pinning potential curvature $\ka$. The main difference between the ferromagnetic domain qubit and the antiferromagnetic one is that the momentum conjugate to the domain wall angle $\Phi$ is the domain wall position $X$ in the former case while it is the angular velocity $\dot\Phi$ in the latter. The fact that the domain wall position is the momentum conjugate to the angle $\Phi$ may pose a problem when coupling qubits in the ferromagnetic scenario. If the qubits are operating in the ``phase regime" (as was assumed in the current proposal), quantum fluctuations in $X$ would be strong in the quantum regime. If two domain wall qubits are coupled by bringing them spatially close together, the two-qubit coupling strength is expected to depend on their spatial separation. Therefore, the fluctuations of the positional variables would lead to fluctuations in the two-qubit coupling strength (which is likely to depend on the inter-qubit distance) and to sources of noise. This issue does not arise in the antiferromagnetic case because the position and the angle variables completely decouple. 

\section*{Acknowledgments}
S.~T. would like to thank Matthieu Dartiailh and Se Kwon Kim for valuable discussions and for the critical reading of the manuscript. This research was supported by CUNY Research Foundation Project \# 90922-07 10.

\end{document}